\documentstyle[preprint,aps]{revtex}
\tightenlines
\begin{document}
\draft
\title{Quantum phase transitions of fractons\\
 ( a fractal scaling theory for the FQHE ) }

\author{Wellington da Cruz}
\address{Departamento de F\'{\i}sica,\\
 Universidade Estadual de Londrina, Caixa Postal 6001,\\
Cep 86051-970 Londrina, PR, Brazil\\
E-mail address: wdacruz@exatas.uel.br}
\date{\today}
\maketitle
\begin{abstract}

We consider the quantum phase transitions of fractons in correspondence 
with the quantum phase transitions of the fractional quantum Hall effect-FQHE. 
We have that the Hall states can be modelled by fractons, known as 
charge-flux systems which satisfy a fractal distribution function 
associated with a fractal von Neumann entropy. In our formulation, 
the universality classes of the fractional quantum Hall transitions, 
are considered as fractal sets of dual topological quantum numbers 
filling factors labelled by the Hausdorff dimension $h$ ( $1 < h < 2$ ) 
of the quantum paths of fractons. In this way we have defined, 
associated to these universality classes, a scaling exponent as $\kappa=1/h$, 
such that when $h$ runs into its interval of definition, we obtain  
$ 1 \gtrsim \kappa \gtrsim 0.5 $. The behavior of this 
scaling exponent, topological in character, is in agreement 
with some experimental values claimed 
in the literature. Thus, according to our approach we have a fractal 
scaling theory for the FQHE which distinguishes diverse 
universality classes for the fractional quantum Hall transitions. 

\end{abstract}

%\narrowtext

\newpage

\section{Introduction}

Quantum phase transitions in complex systems\cite{R1}, like fractional quantum Hall 
effect-FQHE, have been object of great attention in the last fews years 
focussed on a better understanding of 
these two-dimensional strongly correlated electron systems. This type of
transition occurs in a quantum critical point at $T=0$ when 
some parameter of the Hamiltonian is varied. The transition is 
characterized by an order parameter  
which is zero in the disordered phase, and non-zero and non-unique in the ordered phase. 
In that critical point divergences of the correlation lenght and the 
correlation time occur and the observables present power laws 
on the external parameters. In this way some 
exponents characterize the critical behavior of a particular phase transition. 
Now, in the literature, theoretical and experimental analysis of the FQHE 
have considered the Hall transitions in the 
same universality class with scaling exponent 
 $\zeta=0.45\pm 0.05$ and 
a universal critical exponent $\gamma=2.3\pm 0.1$\cite{R2,R3}. On the other hand, 
there is a suggestion that microscopic details can imply in different universality 
classes for the FQHE\cite{R3}. In contrast, taking into account 
only global properties ( as the modular group and the fractal dimension of quantum paths ), 
our fractal scaling theory for the FQHE shows us subtle differences between  
fractal sets of filling factors. This theory signalizes distinct universality classes 
for the fractional quantum Hall phase transitions which in some sense 
is in agreement with other experimental results. As we can verify\cite{R1}, 
quantum phase transitions remain with too much open questions to be investigated. Thus, 
we propose to discuss, according our view, a controversial point on 
the universality class of the FQHE. 
 
The paper is organized as follows, 
in the second section we review our fractal approach to the fractional 
spin particles and we make a connection with the FQHE; in the third section 
we discussed the entanglement content of the Hall states in terms of an 
entanglement measure for the universal classes of fractons and we extract the result that 
Hall states of a given universality class have the same amount 
of entanglement; in the fourth section we introduced our fractal 
scaling theory for the FQHE which gives us a new perspective for that phenomenon; and in the 
last section we finish with some remarks.

\section{Universal classes of fractons}

The mathematical ideas about fractal geometry\cite{R4} have been applied in 
various contexts of physics, as in nonlinear and 
nonequilibrium phenomena\cite{R5}. Here, in this Letter, we show that the FQHE\cite{R6} 
has a fractal-like structure such that the universality classes of the quantum Hall 
transitions constitute fractal sets labelled by the Hausdorff dimension 
defined within the interval $1$$\;$$ < $$\;$$h$$\;$$ <$$\;$$ 2$ and associated with 
fractal curves ( continuous functions but nowhere differentiable ) of 
objects called fractons. These ones are charge-flux systems defined 
in two-dimensional multiply connected space and they carry 
rational or irrational values of spin. The topological meaning 
of the dual filling factors which characterizes the FQHE comes 
from the connection between the fractal dimension of the 
quantum paths of fractons and these quantum numbers. Thus, we have 
introduced {\it a new geometric insight} for understanding that phenomenon
\footnote{Another 
fractal formulation in connection with the FQHE was discussed in\cite{R35,R36}.}.

Topological quantum numbers are insensitive to the imperfections 
of the systems and so the {\it fractal dimension} as a topological invariant 
makes robust the FQHE properties. By contrast, in the literature, some 
authors have considered 
the concepts of Chern numbers and 
Fredholm indices for the integer quantum Hall effect, but for the FQHE 
this research line is an open challenge. The Fredhom index is related to the creation and 
annihilation operators of many-body quantum mechanics and 
in some cases coincides with the Chern number\cite{R7}. The mathematical 
mechanism responsible for the FQHE has been also considered using some techniques of 
noncommutative geometry in connection with twisted higher index theory of elliptic 
operators on orbifold ( its definition generalizes the idea of a manifold ) 
covering spaces of compact good orbifolds, 
where the topological nature of the Hall condutance 
is stable under small deformations of a Hamiltonian, with the interaction 
simulates by the negative curvature of the hyperbolic structure of the model. 
The twisted 
higher index can be a fraction when the orbifold is not smooth and 
the topological invariant is namely the orbifold Euler characteristic\cite{R8}. The 
connection between the FQHE and topological Chern-Simons field theories 
is another route of investigation\cite{R9}.  
Therefore, our fractal approach offers 
{\it a possible alternative} for understanding some deeper 
mathematical and physical features underlying the FQHE
\cite{R10,R11,R12,R13,R14,R15,R16,R17,R18}. 

Fractons satisfy a fractal distribution 
function associated with a fractal von Neumann entropy and they are 
classified in universal classes of particles or quasiparticles
\footnote{ The universal classes of fractons are defined in the same way 
that fermions and bosons constitute universal classes of particles 
with semi-integer and integer values of spin, satisfying the fermionic 
and bosonic distribution functions, respectively. Thus, we have 
universal classes of particles with rational or irrational values 
of spin satisfying a specific fractal distribution function.}. Our 
formulation was introduced in\cite{R10} 
and in particular, we have found an expression which relates 
the fractal dimension $h$ and the spin $s$ of the particles, 
$h=2-2s$, $0 < s < \frac{1}{2}$. This result is  
analogous to the fractal dimension formula 
of the graph of the functions, in the context of the fractal geometry, 
and given by: $\Delta(\Gamma)=2-H$, where $H$ is 
known as H\"older exponent, with $0 < H < 1$\cite{R4}. 
The bounds of the fractal dimension $1$$\;$$ < $$\;$$\Delta(\Gamma)$$\;$$ 
<$$\;$$ 2$ needs to be obeyed in order for a function 
to be a fractal, so the bounds of our parameter $h$ are 
defined such that, for $h=1$ 
we have fermions, for $h=2$ we have bosons and for 
$1$$\;$$ < $$\;$$h$$\;$$ <$$\;$$ 2$ we have fractons. The H\"older exponent 
characterizes irregular functions which appears in diverse physical systems. 
For instance, in the Feynman path integral approach of the quantum mechanics, 
the fractal character of the quantum paths was just observed\cite{R19}.   

The fractal properties of the 
quantum paths can be extracted from the propagators of the particles 
in the momentum space\cite{R10,R20} and so our expression relating 
$h$ and $s$ can once more be justified. On the other hand, the physical formula 
introduced by us, when we consider 
the spin-statistics relation $\nu=2s$, is written as $h=2-\nu$, $0 < \nu < 1$. 
This way, a fractal spectrum was defined taking into account a mirror symmetry:

\begin{eqnarray}
\label{e.100}
h-1&=&1-\nu,\;\;\;\; 0 < \nu < 1;\;\;\;\;\;\;\;\;h-1=
\nu-1,\;\;\;\;\;\;\; 1 <\nu < 2;\nonumber\\
h-1&=&3-\nu,\;\;\;\; 2 < \nu < 3;\;\;\;\;\;\;\;\;
 h-1=\nu-3,\;
\;\;\;\;\;\; 3 <\nu < 4;\;etc.
\end{eqnarray}

The statistical weight for these classes of fractons is given by\cite{R10}

\begin{equation}
\label{e11}
{\cal W}[h,n]=\frac{\left[G+(nG-1)(h-1)\right]!}{[nG]!
\left[G+(nG-1)(h-1)-nG\right]!}
\end{equation}

and from the condition of the entropy be a maximum, we obtain 
the fractal distribution function

\begin{eqnarray}
\label{e.44} 
n[h]=\frac{1}{{\cal{Y}}[\xi]-h}.
\end{eqnarray}

The function ${\cal{Y}}[\xi]$ satisfies the equation 

\begin{eqnarray}
\label{e.4} 
\xi=\biggl\{{\cal{Y}}[\xi]-1\biggr\}^{h-1}
\biggl\{{\cal{Y}}[\xi]-2\biggr\}^{2-h},
\end{eqnarray}

\noindent with $\xi=\exp\left\{(\epsilon-\mu)/KT\right\}$.

We understand the 
fractal distribution function as a quantum-geometrical 
description of the statistical laws of nature, 
since the quantum path is a fractal curve ( this property was noted by Feynman ) and this 
reflects the Heisenberg uncertainty principle. The Eq.(\ref{e.44}) 
embodies nicely this subtle information about the quantum paths 
associated with the particles. 

We can obtain for any class its distribution function considering the
Eqs.(\ref{e.44},\ref{e.4}). For example, 
the universal class $h=\frac{3}{2}$ with distinct values of spin 
$\biggl\{\frac{1}{4},\frac{3}{4},\frac{5}{4},\cdots\biggr\}_{h=\frac{3}{2}}$, 
has a specific fractal distribution

\begin{eqnarray}
n\left[\frac{3}{2}\right]=\frac{1}{\sqrt{\frac{1}{4}+\xi^2}}.
\end{eqnarray}

\noindent This result coincides with another 
one of the literature of fractional spin particles for 
the statistical parameter $\nu=\frac{1}{2}$\cite{R21}, 
however our interpretation is completely distinct. This particular example, 
shows us that the fractal distribution is the same 
for all the particles into the universal class labelled by $h$ and with 
different values of spin.  Thus, we emphasize 
that in our formulation the spin-statistics 
connection is valid for such fractons. The authors in\cite{R21} 
never make mention to this possibility. Therefore, 
our results give another perspective for the fractional 
spin particles or anyons\cite{R9}. On the other hand, we can obtain straightforward the 
Hausdorff dimension associated to the quantum paths of 
the particles with any value of spin.  

We also have
 
\begin{eqnarray}
\xi^{-1}=\biggl\{\Theta[{\cal{Y}}]\biggr\}^{h-2}-
\biggl\{\Theta[{\cal{Y}}]\biggr\}^{h-1},
\end{eqnarray}

\noindent where

\begin{eqnarray}
\Theta[{\cal{Y}}]=
\frac{{\cal{Y}}[\xi]-2}{{\cal{Y}}[\xi]-1}
\end{eqnarray}

\noindent is the single-particle partition function. 
We verify that the classes $h$ satisfy a duality symmetry defined by 
${\tilde{h}}=3-h$. So, fermions and bosons come as dual particles. 
As a consequence, we extract a fractal 
supersymmetry which defines pairs of particles $\left(s,s+\frac{1}{2}\right)$. 
This way, the fractal distribution function appears as 
a natural generalization of the fermionic and bosonic 
distributions for particles with braiding properties. Therefore, 
our approach is a unified formulation 
in terms of the statistics which each universal class of 
particles satisfies, from a unique expression 
we can take out any distribution function. In some sense , we can say that 
fermions are fractons of the class $h=1$ and  
bosons are fractons of the class $h=2$.

The free energy for particles in a given quantum state is expressed as

\begin{eqnarray}
{\cal{F}}[h]=KT\ln\Theta[{\cal{Y}}].
\end{eqnarray}

\noindent The fractal von Neumann entropy per state in terms of the 
average occupation number is given as\cite{R10,R11} 

\begin{eqnarray}
\label{e5}
{\cal{S}}_{G}[h,n]&=& K\left[\left[1+(h-1)n\right]\ln\left\{\frac{1+(h-1)n}{n}\right\}
-\left[1+(h-2)n\right]\ln\left\{\frac{1+(h-2)n}{n}\right\}\right]
\end{eqnarray}

\noindent and it is associated with the fractal distribution function Eq.(\ref{e.44}).

Now, as we can check, each universal class $h$ of particles, 
within the interval of definition has its entropy defined 
by the Eq.(\ref{e5}). Thus, for fractons of the self-dual class
$\biggl\{\frac{1}{4},
\frac{3}{4},\frac{5}{4},\cdots\biggr\}_{h=\frac{3}{2}}$, we obtain
  
\begin{eqnarray}
{\cal{S}}_{G}\left[\frac{3}{2}\right]=K\left\{(2+n)\ln\sqrt{\frac{2+n}{2n}}
-(2-n)\ln\sqrt{\frac{2-n}{2n}}\right\}. 
\end{eqnarray}

\noindent The Fermi and Bose statistics ( and the respective entropies, free energies ) associated to 
the universal classes of the fermions and bosons are recuperated promptly.

We have also introduced the topological concept of fractal index, 
which is associated with each class. As we saw, $h$ is a geometrical parameter 
related to the quantum paths of the particles and so, we define\cite{R12} 

\begin{equation}
\label{e.1}
i_{f}[h]=\frac{6}{\pi^2}\int_{\infty(T=0)}^{1(T=\infty)}
\frac{d\xi}{\xi}\ln\left\{\Theta[\cal{Y}(\xi)]\right\}.
\end{equation}

\noindent For the interval of the definition $ 1$$\;$$ \leq $$\;$$h$$\;$$ \leq $$\;$$ 2$, there 
exists the correspondence $0.5$$\;$$ 
\leq $$\;$$i_{f}[h]$$\;$$ \leq $$\;$$ 1$, which signalizes 
a connection between fractons and quasiparticles of the conformal field theories, 
in accordance with the unitary $c$$\;$$ <$$\;$$ 1$ 
representations of the central charge. Thus, we have established 
a connection between fractal geometry and 
number theory, given that the dilogarithm function appears 
in this context, besides another branches of mathematics\cite{R22}. 

Such ideas can be applied in the context of the FQHE. This phenomenon is 
characterized  by the filling factor parameter $f$, and for 
each value of $f$ we have the 
quantization of Hall resistance and a superconducting state 
along the longitudinal direction of a planar system of electrons, which are
manifested by semiconductor doped materials, i.e., heterojunctions 
under intense perpendicular magnetic fields and lower 
temperatures\cite{R6}. 

The parameter $f$ is defined by $f=N\frac{\phi_{0}}{\phi}$, where 
$N$ is the electron number, 
$\phi_{0}$ is the quantum unit of flux and
$\phi$ is the flux of the external magnetic field throughout the sample. 
The spin-statistics relation is given by 
$\nu=2s=2\frac{\phi\prime}{\phi_{0}}$, where 
$\phi\prime$  is the flux associated with the charge-flux 
system which defines the fracton $(h,\nu)$. According to our approach 
there is a correspondence between $f$ and $\nu$, numerically $f=\nu$. 
This way, we verify that the filling factors 
experimentally observed  appear into the classes $h$ and from the definition of duality 
between the equivalence classes, we note that the FQHE occurs in pairs 
 of these dual topological quantum numbers.

\section{Entanglement fractal von Neumann entropy}

In\cite{R23} we have defined an entanglement measure for 
the universal classes of fractons in terms of the probability 
distribution $p$ as:

\begin{equation}
\label{e.43}
{\cal{E}}[h,p]=\frac{1}{1-(h-1)p}\left\{-p\log_{2}p -(1-p)\log_{2}(1-p)\right\},
\end{equation}

\noindent where $0 \leq p\leq 1$, is the probability of the  system to 
be in a microstate with 
entanglement between ocuppation-numbers of the modes 
considered empty, partially or completely filled.

The entanglement properties of fractons ( Hall states ) 
can be analyzed considering the Eq.(\ref{e.43}) and we verify that the 
entanglement increases 
when we run in the interval  $1 < h < 2$, for instance, 
${\cal{E}}[h=4/3]< {\cal{E}}[h=3/2]<
{\cal{E}}[h=5/3]$. Given the sequence,

\begin{eqnarray}
&&\cdots\rightarrow\biggl\{\frac{2}{3},\frac{4}{3},\frac{8}{3},
\cdots\biggr\}_{h=\frac{4}{3}}\rightarrow\;\;
\biggl\{\frac{1}{2},\frac{3}{2},\frac{5}{2},
\cdots\biggr\}_{h=\frac{3}{2}}\rightarrow
\biggl\{\frac{1}{3},\frac{5}{3},\frac{7}{3},
\cdots\biggr\}_{h=\frac{5}{3}}\rightarrow\cdots,\nonumber
\end{eqnarray}

\noindent and the entanglement measure in 
terms of the filling factors

\begin{eqnarray}
\label{e12}
{\cal{E}}[2-\nu,p]&=&\frac{1}{1-(1-\nu)p}\left\{-p\log_{2}p -(1-p)\log_{2}(1-p)\right\}, 
\end{eqnarray}

\noindent with $0 < \nu < 1$, we obtain ${\cal{E}}[\nu=2/3]< {\cal{E}}[\nu=1/2]<
{\cal{E}}[\nu=1/3]$. For the other members of the classes we need to consider the 
fractal spectrum Eq.(\ref{e.100}). This way, we verify that the 
Eq.(\ref{e.43}) for the class $h$, is 
the same for all the members of the class and so, in terms of their entanglement content, 
different Hall states are equivalent. The understanding that something in this sense 
can be provided by a quantitative theory of entanglement for 
complex quantum systems was envisaged by Osborne-Nielsen 
in\cite{R24}. Therefore, we have obtained  
a result, in the context of the FQHE, which just realizes this perception. Observe that our approach 
gives information about the entanglement for any possible wave function 
associated with a specific value of the filling factor. In another route we can consider 
the LLL for fractons, i.e. if the temperature 
 is sufficiently low and $\epsilon <\mu$, we can check 
 that the mean ocuppation number Eq.(\ref{e.44}) is given 
 by $n=\frac{1}{2-h}$, and so the fractal parameter $h$ 
 regulates the number of particles in each quantum state. At $T=0$ and $\epsilon> \mu$, $n=0$
 if $\epsilon > \epsilon_{F}$ and $n=\frac{1}{2-h}$ 
 if $\epsilon < \epsilon_{F}$, hence we get a step distribution, 
 taking into account the Fermi energy $\epsilon_{F}$ and $h\neq 2$. We can check that for 
 $h=\frac{4}{3},\frac{3}{2},\frac{5}{3}$ we obtain $n=\frac{3}{2},\frac{2}{1},\frac{3}{1}$, 
 respectively.  In the first case we have three particles for two states, in the second 
 case two particles for one state and in the last case three particles for one state. So when 
 we run in the interval $ 1 < h < 2$ we gain more particles for each possible state. In 
 some sense fractons can be understood as {\it quasifermions} when near 
 the universal class $h=1$ and as $quasibosons$ when near the universal class $h=2$. 
 The entanglement of the FQHE increases because we have more particles ( fractons ) 
 and less states. On the other hand, in terms of the filling factors, 
 the average ocuppation number can be written as
 $n=\frac{1}{\nu}$, $0 < \nu < 1$;  $n=\frac{1}{2-\nu}$, $1 < \nu < 2$; $n=\frac{1}{\nu-2}$, 
 $2 < \nu < 3$; etc. The behavior of the step distribution confirmes 
our former analysis: the ground state of the FQHE is a stronger entangled state and the 
entanglement between ocuppation-numbers of fractons in the LLL and the rest of 
the system shows us quantum correlations which can be quantified.

All these results agree with the entanglement properties of the Laughlin 
wave functions and those generated by the K-matrix\cite{R25}. On the other hand, 
the FQHE understood in terms of the composite fermions or composite 
bosons are non-entangled as observed in\cite{R26}, so in contrast, fractons appear 
as a suitable system for study the quantum correlations of the FQHE. 
Thus the suggestion that ideas of the quantum information science can give insights for 
understanding some complex quantum systems\cite{R24} is manifested in 
our definition of entanglement measure for the universal classes of fractons. As we saw, 
the entanglement increases in the interval $ 1 < h < 2 $ and this suggests fracton 
qubits as a physical resource for quantum computing. In the literature FQHE qubits 
associated with the geometrical characteristic of the fractional spin particles 
have been exploited\cite{R27}. The quantum Hall phase transitions discussed by us 
were obtained considering global properties as the modular symmetry and 
the Hausdorff dimension associated to the quantum paths of the particles, 
so some peculiarities of the FQHE, in particular, do not depend on the microscopic 
details of this strongly interacting system\cite{R10}.

Here, we observe that our approach, in terms of fractal sets of 
dual topological filling factors, embodies 
the structure of the modular group as discussed in the literature
\cite{R18,R28} and the quantum Hall transitions satisfy some 
properties related with the Farey sequences of rational numbers. The 
transitions allowed are those generated by the condition
 $\mid p_{2}q_{1}
-p_{1}q_{2}\mid=1$, 
with $\nu_{1}=\frac{p_{1}}{q_{1}}$ and $\nu_{2}=
\frac{p_{2}}{q_{2}}$. In\cite{R28} the properties of integer and fractional 
quantum Hall effect were considered in terms of a subgroup of the modular group 
$SL(2,{\bf Z})$, such that the group acts on the upper-half complex plane 
parameterised by a complex conductivity 
and so generates the phase diagram of the quantum Hall effect. The rules 
obtained there are the same that our formulation\cite{R10}, but we have defined 
universality classes of the quantum Hall 
transitions in terms of fractal sets labelled by a fractal parameter. On the other one, 
our theoretical description consider the Hall 
states modelled by systems of fractons with a specific value of spin.

\section{Fractal scaling theory for the FQHE transitions}

As we discussed earlier, the universality classes of the fractional quantum Hall transitions 
have been considered as fractal sets of dual topological quantum numbers 
filling factors labelled by the Hausdorff dimension. Such approach has the fractal 
parameter $h$ as a critical exponent for the correlation length associated 
with a scaling exponent defined as $\kappa=1/h$. So, the members ( possible 
Hall states characterized by the values of the filling factor ) 
of a given class, for example, $\left\{\frac{1}{3},\frac{5}{3},\cdots\right\}_{h=\frac{5}{3}}
$or $\left\{\frac{2}{3},\frac{4}{3},\cdots\right\}_{h=\frac{4}{3}}$, 
share the same scaling exponent $\kappa=0.6$ and $\kappa=0.75$, respectively. 
We can establish families of quantum Hall phase transitions as

\begin{eqnarray}
&&\biggl\{\frac{1}{5},\frac{9}{5},\frac{11}{5},\frac{19}{5},
\cdots\biggr\}_{h=\frac{9}{5}}^{\kappa=0.55}\rightarrow\;\;\;
\biggl\{\frac{2}{9},\frac{16}{9},\frac{20}{9},\frac{34}{9},
\cdots\biggr\}_{h=\frac{16}{9}}^{\kappa=0.56}\rightarrow
\biggl\{\frac{1}{4},\frac{7}{4},\frac{9}{4},\frac{15}{4},
\cdots\biggr\}_{h=\frac{7}{4}}^{\kappa=0.57}\rightarrow\nonumber\\ 
&&\biggl\{\frac{2}{7},\frac{12}{7},\frac{16}{7},\frac{26}{7},
\cdots\biggr\}_{h=\frac{12}{7}}^{\kappa=0.58}\rightarrow\; 
\biggl\{\frac{1}{3},\frac{5}{3},\frac{7}{3},\frac{11}{3},
\cdots\biggr\}_{h=\frac{5}{3}}^{\kappa=0.60}\rightarrow \;\;\;\;
\biggl\{\frac{2}{5},\frac{8}{5},\frac{12}{5},\frac{18}{5},
\cdots\biggr\}_{h=\frac{8}{5}}^{\kappa=0.625}\rightarrow\nonumber\\
 &&\biggl\{\frac{3}{7},\frac{11}{7},\frac{17}{7},\frac{25}{7},
\cdots\biggr\}_{h=\frac{11}{7}}^{\kappa=0.63}\rightarrow\;
 \biggl\{\frac{1}{2},\frac{3}{2},\frac{5}{2},\frac{7}{2},
\cdots\biggr\}_{h=\frac{3}{2}}^{\kappa=0.66}
\rightarrow\;\;\;\;\;\biggl\{\frac{4}{7},\frac{10}{7},\frac{18}{7},\frac{24}{7},
\cdots\biggr\}_{h=\frac{10}{7}}^{\kappa=0.70}\rightarrow\nonumber\\ 
&&\biggl\{\frac{3}{5},\frac{7}{5},\frac{13}{5},\frac{17}{5},
\cdots\biggr\}_{h=\frac{7}{5}}^{\kappa=0.71}\rightarrow\;\;\;
\biggl\{\frac{2}{3},\frac{4}{3},\frac{8}{3},\frac{10}{3},
\cdots\biggr\}_{h=\frac{4}{3}}^{\kappa=0.75}
\rightarrow\;\;\;\biggl\{\frac{5}{7},\frac{9}{7},\frac{21}{7},\frac{23}{7},
\cdots\biggr\}_{h=\frac{9}{7}}^{\kappa=0.77}
\rightarrow\nonumber\\ 
&&\biggl\{\frac{3}{4},\frac{5}{4},\frac{11}{4},\frac{13}{4},
\cdots\biggr\}_{h=\frac{5}{4}}^{\kappa=0.80}\rightarrow\;\;\;
\biggl\{\frac{7}{9},\frac{11}{9},\frac{25}{9},
\frac{29}{9},\cdots\biggr\}_{h=\frac{11}{9}}^{\kappa=0.81}
\rightarrow\biggl\{\frac{4}{5},\frac{6}{5},\frac{14}{5},\frac{16}{5},
\cdots\biggr\}_{h=\frac{6}{5}}^{\kappa=0.82}.\nonumber
\end{eqnarray}

\noindent We can see, in this sequence, that we have sets with values of scaling 
exponent $\kappa\sim 0.5, 0.6, 0.7, 0.8$. So, up to now, the experiments did not detect 
these subtle differences between distinct universality classes for the FQHE. 
When $h$ runs into the interval of definition, the scaling exponent stays into 
$ 1 \gtrsim \kappa \gtrsim  0.5$. On the other hand, in the literature, values 
for the scaling exponent $\zeta\sim 0.45, 0.6, 0.77$ have been reported, 
with $\zeta=1/z\gamma$, 
 where $z$ is a dynamical critical exponent and $\gamma$ is a universal critical exponent. 
 We observe that for adjacent quantum Hall phase transitions, the scaling exponents 
 do not present appreciable difference between them. So, in the experiments, 
 this is not detected and 
 the scaling exponent appears with the same value for all the filling factors. 
 {\it We believe that this is not the case}. We have here fine differences between distinct 
 classes of Hall states. In\cite{R29} was verified a transition from the 
 integer filling factor 
 $\nu=1$  to the fractional filling factor $\nu=2/3$ 
 with scaling exponent $\zeta=0.77\pm 0.02$ which coincides with 
 our value of $\kappa=0.75$ just for the class $h=4/3$ where the filling 
 factor considered appears. We enphasize that all the members of this class 
 have the same scaling exponent. In this sense, these Hall states 
 are in the same universality class. However, the transitions $3/5-2/3$, $1-4/3$, $5/3-2$ 
 have been reported with scaling behavior $\zeta=0.45\pm 0.05$ and for the transition $2/3-1$, 
$\zeta=0.77$\cite{R30}. Also, for the transition $1/3$ to $2/5$, was 
found\cite{R31} the value of $\zeta=0.43\pm 0.02$ while via our approach we have that 
for the filling factor $1/3$ we obtain $\kappa=0.6$ and for $\nu=2/5$, $\kappa=0.625$. 
In another experiment, a scaling exponent $\zeta=0.5\pm 0.1$ 
was found which is independent of material, density, mobility, experimental 
technique, temperature and the filling factor\cite{R32}. As we can see, our understanding 
of the quantum phase transitions, in this quantum complex system-FQHE, until now 
is not complete and controversial\cite{R33}. Thus, our fractal scaling theory 
signalizes some correspondence between the scaling exponent, $\kappa$, 
for quantum phase transitions of fractons and the scaling exponent, $\zeta$, 
for fractional quantum Hall transitions. This gives us the opportunity of to distinguish 
universality classes for the FQHE and so we suggest to the experimentalists looking for 
this fractal scaling behavior in these two-dimensional strongly correlated electron systems.

Here we emphasize the topological character of the 
scaling exponent $\kappa$, because it is derived of a geometrical parameter associated 
to the fractal curves of the quantum particles. As this reflects the Heisenberg 
uncertainty principle
\footnote{{Quantum fluctuations are driven by the Heisenberg uncertainty 
principle and this one expresses the fractal character of the quantum paths.}}, 
quantum fluctuations are encoded in this concept of Hausdorff dimension 
associated to the quantum paths of fractons. Also, as the filling factors are topological quantum 
numbers in principle and this characteristic in our theoretical formulation 
is extracted from the connection between $h$ and $\nu$, via the fractal spectrum 
Eq.(\ref{e.100}), we can think $\kappa$ as a signature of some kind of topological phase transition 
between fractal sets of filling factors and so they are in correspondence with the quantum 
Hall phase transitions.

\section{Conclusions}

We have suggested in this article that the 
fractional quantum Hall phase transitions obey a fractal scaling theory 
which implies distinct universality classes for that phenomenon. According to our 
approach these transitions are characterized by a scaling exponent, $\kappa$, which 
is related with the fractal dimension, $h$, of the quantum paths of fractons. 
In this way the topological character of the scaling exponent signalizes transitions
between fractal sets of filling factors which do not depend on the microscopic details of 
the physical system. These topological quantum phase transitions show, 
for values of the scaling exponent into the interval $ 1 \gtrsim \kappa \gtrsim 0.5 $, 
correspondence with some experimental values. Besides these robust aspects, 
our approach via the concept 
of the entanglement measure for the universal classes of fractons has extracted 
the result that the Hall states within of the same universality class 
have equivalent entanglement. Also along of our discussion 
is implicit that our approach to the FQHE 
reproduces all experimental data and can predicting 
the occurrence of this phenomenon 
for other filling factors. These topological quantum numbers are obtained from first-principles 
and the quantum Hall states are modelled  by fractons which carry rational 
or irrational values of spin and satisfy a fractal distribution function 
associated with a fractal von Neumann entropy. Therefore, the physical 
scenario takes place and the thermodynamical properties 
of such systems can be investigated.

We emphasize that our approach is supported by symmetry principles such as 
mirror symmetry behind the fractal spectrum, duality symmetry 
between the universal classes of particles, fractal supersymmetry, modular group behind 
the universality classes of the quantum Hall transitions. Besides 
these stronger arguments given that the symmetry groups have 
great importance for understanding the physical theories even 
before any dynamics, we have established a connection between 
physics, fractal geometry and number theory. Along this discussion we 
have obtained fractal sets of dual 
topological quantum numbers filling factors associated with the FQHE. 
The universality classes of the 
quantum Hall transitions were established and the fractal 
geometry of nature in this context is manifest
in a simple and intuitive way.

Here, we observe a possible connection between ours 
results and a discussion in\cite{R34}, which consider if the universality 
classes of a phase transition 
depends on the fractional statistics. For that, we have an 
affirmative answer according to our fractal approach to the fractional spin particles. 
Also, as the fractons obey the spin-statistics connection and we have verified 
a fractal supersymmetry between the universal classes of these objects, 
we stay with the 
intuition that we can advance a hypothesis on the analyticity of a given amplitude 
introduced in the context of some models studied in\cite{R34}.  

Finally, we believe that our fractal scaling theory contains some ideas 
which can be considered within the structure of some other theory ahead, such that be 
possible a more deeper understanding of the FQHE.

\end{document}